\documentclass{kapproc} 

\usepackage[dvips]{graphicx}
\upperandlowercase

\kluwerbib 

\begin{document}

\articletitle{Modelling the Red Halos of Blue Compact Galaxies}

\articlesubtitle{}

\author{Erik Zackrisson, Nils Bergvall, Thomas Marquart, Lars Mattsson}
\affil{Department of Astronomy \& Space Physics\\
Box 515\\
SE-751 20 Uppsala\\
Sweden}
\author{G\"{o}ran \"{O}stlin}
\affil{Department of Astronomy\\
SE-106 91 Stockholm\\
Sweden}

\begin{abstract}
Optical/near-IR broadband photometry of the halos of blue compact galaxies (BCGs) have revealed a very red spectral energy distribution, which cannot easily be reconciled with any normal, metal-poor stellar population. Here, three possible explanations for the red excess are explored: nebular emission, metal-rich stars and a stellar population with a very bottom-heavy initial mass function (IMF). We find, that nebular emission in BCG halos would produce very blue near-IR colours, and hence fails to explain the observed red excess. Although metal-rich stars may in principle explain the colours observed, the required stellar metallicity is very high (solar or higher), which would be a curious halo property given the low gas metallicity ($\sim10\%$ solar) of the central starburst in these objects. A stellar population with a low to intermediate metallicity and a very bottom-heavy IMF does however adequately reproduce the observed BCG halo colours. A bottom-heavy IMF also proves successful in explaining a similar red excess observed in the halos of stacked edge-on disk galaxies from the Sloan Digital Sky Survey (SDSS). This may indicate that halos dominated by low-mass stars is a phenomenon common to galaxies of very different Hubble types.
\end{abstract}
\section*{Introduction}
Optical/near-IR broadband photometry of the faint halos surrounding BCGs have revealed a very red spectral energy distribution (Bergvall \& \"Ostlin \cite{Bergvall & Östlin}; Bergvall et al. \cite{Bergvall et al.}; \"Ostlin et al., in preparation), which cannot easily be reconciled with a metal-poor stellar population like that in the halo of the Milky Way. The red excess is not likely to be caused by dust reddening given the upper limits on the presence of cold dust in BCG halos inferred from ISO data (Bergvall et al., in preparation), and the low extinction measured by the H$\alpha$/H$\beta$ Balmer decrement in the central starburst. In the case of Haro 11, the BCG with the reddest halo observed so far, ISO observations furthermore rule out near-IR emission from warm dust as an explanation for the red colours (Bergvall \& \"Ostlin \cite{Bergvall & Östlin}).

Recently, a somewhat similar red excess was also noted in halos detected around edge-on disk galaxies in stacked optical data from the SDSS (Zibetti, White \& Brinkmann \cite{Zibetti et al.}). Are these red halo phenomena related, and if so -- what is the origin of the red excess? Here, models of spectral evolution are used to test three different possible explanations for the red halo colours: metal-rich stars, nebular emission and a stellar population with a very bottom-heavy IMF. 

\section{Metal-Rich Stars}
\begin{figure}[t]
\centering
\includegraphics[scale=0.5]{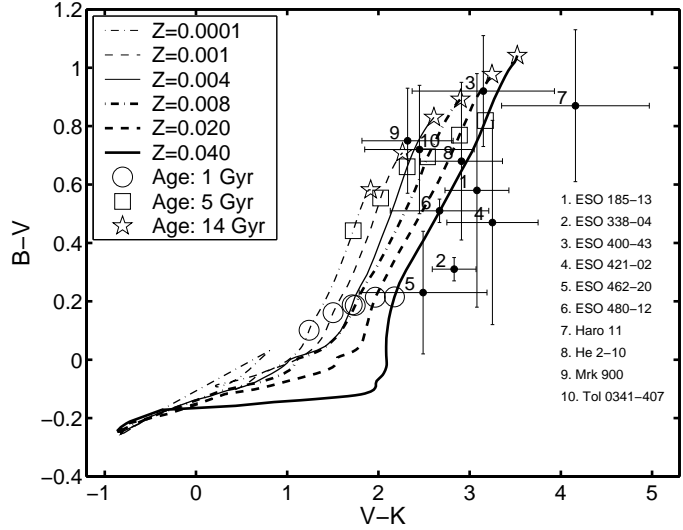}
\caption{Observed colours of BCG halos (crosses indicating $1\sigma$ error bars), compared to the predictions of P\'EGASE.2 (lines) for stellar populations with a Salpeter IMF and an exponentially declining star formation rate (SFR$(t)\propto \exp{-t/\tau}$) with $\tau=1$ Gyr. The different lines correspond to constant metallicities of $Z=0.0001$ (thin dash-dotted), $Z=0.001$ (thin dashed), $Z=0.004$ (thin solid), $Z=0.008$ (thick dash-dotted), $Z=0.020$ (thick dashed) and $Z=0.040$ (thick solid). Markers along the evolutionary sequences indicate population ages of 1 Gyr (circle), 5 Gyr (square) and 14 Gyr (pentagram). For many of the halos, only the highest metallicities ($Z\geq 0.020$) provide a reasonable fit.}
\label{highzfig}
\end{figure}

In Bergvall \& \"Ostlin (\cite{Bergvall & Östlin}), metal-rich stellar populations were suggested as an explanation for the red excess of BCG halos. In Fig.~\ref{highzfig}, we use the P\'EGASE.2 model of spectral evolution (Fioc \& Rocca-Volmerange \cite{Fioc & Rocca-Volmerange}) to show that although this solution does reasonably well when confronted with the extended BCG halo data set of Bergvall et al. (\cite{Bergvall et al.}), the metallicities of many of the halos would have to be very high (solar or higher). Other spectral evolutionary models (e.g. Zackrisson et al. \cite{Zackrisson et al.}; Bruzual \& Charlot \cite{Bruzual & Charlot}) confirm this result. Such high stellar metallicities would be very strange given the low metallicity ($\sim$10\% solar) of the gas in the central starburst of these objects. Furthermore, this high-metallicity solution fails to explain the halos of edge-on spirals in the SDSS, as demonstrated in Fig.~\ref{imffig_SDSS} (dash-dotted lines).

\section{Nebular Emission}
In principle, the colours of BCG halos could be affected by nebular emission originating in an extended envelope of gas ionized by hot stars in the central starburst. Since photoionization models predict the spectrum of a complete Str\"omgren sphere to be very blue, this explanation for the red excess was dismissed by Bergvall \& \"Ostlin (\cite{Bergvall & Östlin}). Current halo observations do however not probe the entire Str\"omgren sphere, but only lines of sight through the outer parts of the ionized cloud. Could it be that these regions have a substantially different colour signature? 

\begin{figure}[t]
\centering
\includegraphics[scale=0.5]{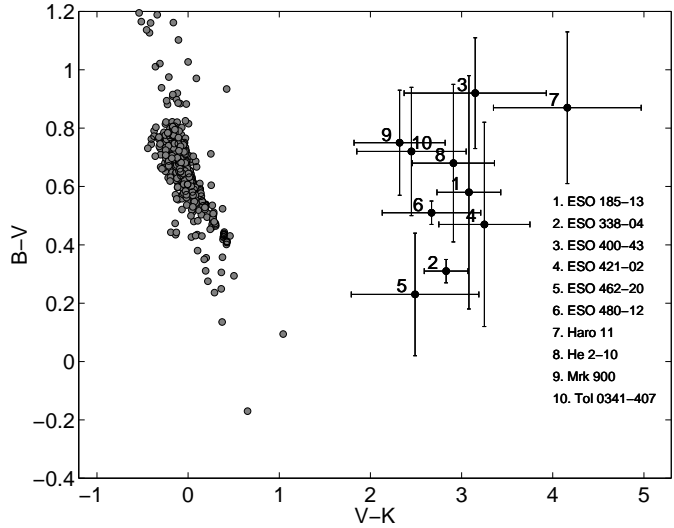}
\caption{Observed colours of BCG halos (crosses indicating $1\sigma$ error bars), compared to the colours predicted for various lines of sight through photoionized nebulae (grey dots). These predictions indicate that nebular emission is much too blue in $V-K$ to explain the observed halo colours.}
\label{nebemfig}
\end{figure}

To explore this possibility, the photoionization model Cloudy (Ferland \cite{Ferland}) has been used to predict the optical/near-IR colours for non-central lines of sight through spherical nebulae. As ionizing continua, representing the stellar population of the central starburst, synthetic spectra for metal-poor ($Z=0.001$) stellar populations with ages of 1 Myr, 10 Myr and 100 Myr were generated using the Zackrisson et al. (\cite{Zackrisson et al.}) spectral evolutionary code. For each stellar population continuum, nebulae with a wide range of density profiles and filling factors were generated. Every resulting model nebula was then recomputed ten times, each time truncated at progressively smaller radii. From these results, the spectral energy distribution of each spherical shell was derived. The spectra of the nebular shells were finally weighted together according to their relative volume along each of the possible lines of sight through the ionized cloud. In total, this resulted in 810 different line-of-sight models, the colours of which are plotted in Fig.~\ref{nebemfig}. Despite a substantial scatter, this more sophisticated treatment of nebular emission still predicts colours much too blue to explain the red excess of BCG halos. Correcting the observed halo colours for potential contamination by nebular emission would therefore only make the red excess even more severe.

\section{A Bottom-heavy Initial Mass Function}
\begin{figure}[t]
\centering
\includegraphics[scale=0.5]{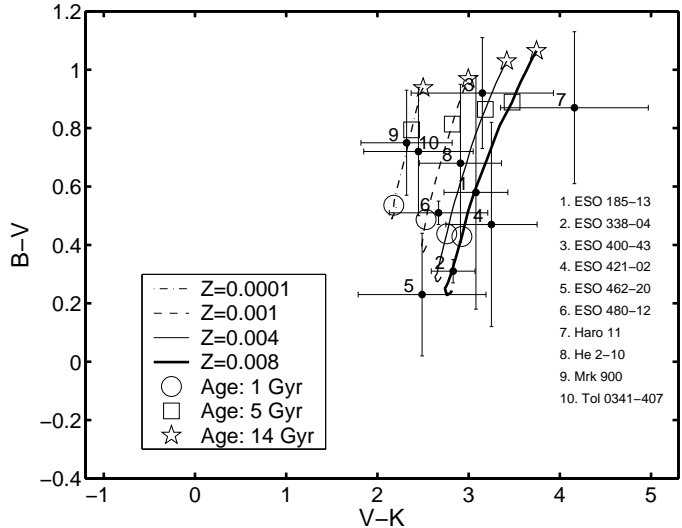}
\caption{Observed colours of BCG halos (crosses indicating $1\sigma$ error bars), compared to the predictions of P\'EGASE.2 (lines) for stellar populations with a bottom heavy IMF ($dN/dM\propto M^{-\alpha}$ with $\alpha=4.50$ for $M=0.08$--120 $M_\odot$) and the same star formation history as in Fig.~\ref{highzfig}. The different lines correspond to constant metallicities of $Z=0.0001$ (thin dash-dotted), $Z=0.001$ (thin dashed), $Z=0.004$ (thin solid), $Z=0.008$ (thick solid). With such an extreme IMF, stellar populations with low to intermediate metallicities ($Z=0.001$--0.008) give a reasonable fit.}
\label{imffig_BVK}
\end{figure}

Could it be that the red excess of BCG halos originates in a stellar population with a peculiar IMF? The most obvious way to produce a very red spectral energy distribution would be to increase the fraction of cool, low-mass stars by adopting a bottom-heavy IMF ($dN/dM\propto M^{-\alpha}$ with $\alpha>2.35$, where $\alpha=2.35$ represents the Salpeter slope). This possibility was briefly explored in Bergvall et al. (\cite{Bergvall et al.}), and found not to significantly improve the situation. However, only IMF slopes of $\alpha\leq 3.35$ were considered. In Fig.~\ref{imffig_BVK}, we use the P\'EGASE.2 model to demonstrate that stellar populations with an extremely bottom-heavy IMF ($dN/dM\propto M^{-\alpha}$ with $\alpha=4.50$, $M=0.08$--$120 \ M_\odot$) in fact can explain the BCG halo colours with low to intermediate stellar metallicities ($Z=0.001$--0.008). 

\begin{figure}[t]
\centering
\includegraphics[scale=0.5]{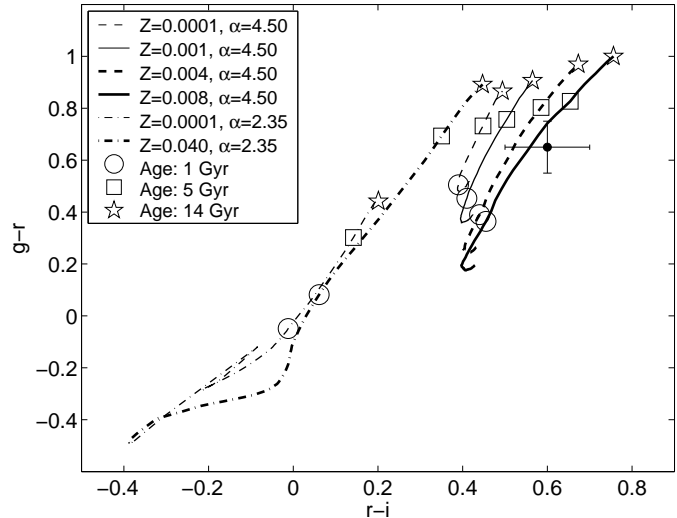}
\caption{The observed colour of the halo detected in stacked SDSS edge-on disk galaxy data (cross indicating $1\sigma$ error bar) compared to the predictions of P\'EGASE.2 (lines) for stellar populations with various metallicities and IMF slopes, $\alpha$, as listed in the legend. The same star formation history as in Fig.~\ref{highzfig} has been adopted. The colours of the SDSS halo is well reproduced by an intermediate-metallicity population ($Z=0.004$--0.008; thick dashed, thick solid) with a bottom-heavy IMF ($\alpha=4.50$, $M=0.08$--120 $M_\odot$). By contrast, populations with Salpeter IMFs (dash-dotted lines) fail to reproduce the observed colours, regardless of metallicity.}
\label{imffig_SDSS}
\end{figure}

Interestingly, stellar populations with similar ages, metallicities and the same bottom-heavy IMF ($\alpha=4.50$) also succeed in explaining the halo detected in stacked data edge-on disk galaxy data from the SDSS (Fig.~\ref{imffig_SDSS}). This conclusion is confirmed by the Zackrisson et al. (\cite{Zackrisson et al.}) spectral evolutionary model as well. A metal-rich stellar population with a Salpeter IMF (thick dashed-dotted line in Fig.~\ref{imffig_SDSS}) on the other hand fails to explain the SDSS halo detection.

\section{Summary}
We find, that nebular emission fails to explain the red excess observed in BCG halos. A population of old, metal-rich stars may in principle explain the BCG halo colours, but only for an extremely high stellar metallicity (solar or higher), which would be very curious given the low metallicity ($\sim10\%$ solar) of the central BCG starburst. Stellar populations with low to intermediate metallicities can however explain the observations, provided that the halo IMF is very bottom-heavy. The latter solution also provides an explanation for a similar red excess observed in the halo detected in stacked edge-on disk galaxy data from the SDSS. This suggests that red halos dominated by low-mass stars may be a phenomenon common to galaxies of very different types. Various tests of this possibility are currently being developed.

\end{document}